\documentclass[%
 reprint,
 amsmath,amssymb,
 aps,
]{revtex4-1}

\pdfoutput=1 

\usepackage{graphicx}
\usepackage{dcolumn}
\usepackage{bm}

\usepackage{color} 

\begin{document}

\preprint{APS/123-QED} 

\title{Non-minimum phase viscoelastic properties of soft biological tissues}

\author{Yo Kobayashi}
\email{yo.kobayashi@me.es.osaka-u.ac.jp}%
\affiliation{Graduate School of Engineering Science, Osaka University, Osaka, Japan\\ and JST-PRESTO \\}%

\author{Naomi Okamura}%
\author{Mariko Tsukune}%
\author{Masakatsu G. Fujie}%
\affiliation{%
 Faculty of Science and Engineering / Future Robotics Organization, Waseda University, Tokyo, Japan 
}%

\author{Masao Tanaka}
\affiliation{Graduate School of Engineering Science, Osaka University, Osaka, Japan}%

\date{\today}


\begin{abstract}

Understanding the visocoelastic properties of soft biological tissues is important for progress in the field of human healthcare. This study analyzes the viscoelastic properties of soft biological tissues using a fractional dynamics model. We conducted a dynamic viscoelastic test on several porcine samples, namely liver, breast, and skeletal muscle tissues, using a plate--plate rheometer. We found that some soft biological tissues have non-minimum phase properties; that is, the relationship between compliance and phase delay is not uniquely related to the non-integer derivative order in the fractional dynamics model. The experimental results show that the actual phase delay is larger than that estimated from compliance. We propose a fractional dynamics model with the fractional Hilbert transform to represent these non-minimum phase properties. The model and experimental results were highly correlated in terms of compliance and phase diagrams and complex mechanical impedance. We also show that the amount of additional phase delay, defined as the increase in actual phase delay compared to that estimated from compliance, differs with tissue type.

\end{abstract}

\maketitle

\section{\label{sec:introduction}Introduction}
\subsection{\label{sec:background}Background}

Understanding the physical phenomena in the human body is important in bioscience and bioengineering. Knowledge of the mechanical properties of human tissues will lead to progress in healthcare. In particular, understanding the viscoelastic properties of biological tissues is key because they can reveal tissue function. These properties are also important for medical treatment because they are closely related to tissue type and disease.

Nevertheless, methods for analyzing the viscoelastic properties of soft biological tissues are not well established. The properties of soft biological tissues are different from those of synthetic materials and thus cannot be directly modeled in the same manner \cite{fung1981biomechanics, fung2013biomechanics}.

The motivation behind this study is to determine the viscoelastic properties of soft biological tissues by modeling their macroscopic properties. Ideally, a model should be strongly correlated with the experimental data and have a small number of parameters. A small number of model parameters is important for determining the viscoelasticity of soft biological tissues, the identification of tissue function, and the robust discrimination of tissue type based on viscoelasticity.

\subsection{\label{relat}Related research}

Many studies have reported that soft biological tissues have viscoelastic properties \cite{ fung1981biomechanics, fung2013biomechanics, maurel1998biomechanical}. An ordinary differential equation, such as that in the Voigt, Maxwell, or Kelvin model, is generally used to model viscoelastic properties\cite{ fung1981biomechanics, fung2013biomechanics,maurel1998biomechanical,wineman2009nonlinear}. Models with a small-order ordinary differential equation do not well fit experimental data for biological tissues. A large-order ordinary differential equation such as that in the generalized Maxwell model can be used to increase model accuracy at the cost of a large number of model parameters. For example, studies have modeled the nonlinear viscoelasticity of the brain \cite{Darvish2001, Miller2002}, kidney \cite{Kim2005}, breast \cite{QIU2018}, liver \cite{Kim2005, Marchesseau2010, Ahn2010,samur2007robotic, Asbach2008, schwartz2005modelling}, skeletal muscle \cite{Wheatley2016}, and subcutaneous tissue \cite{Panda2018}. 

Fractional differential equations have recently been shown to be efficient in modeling the viscoelastic properties of biological tissues. The fractional dynamics model represents a power law response, which is obtained from experimental data of soft biological tissues, with a relatively small number of parameters \cite{Craiem2006,Craiem2010}. For example, the fractional dynamics model was used to model the viscoelastic properties of the lung \cite{Suki1994,Yuan1997,Yuan2000}, the brain \cite{Sack2009,Sack2013}, skeletal muscle \cite{Klatt2010, Sack2013,GRAHOVAC2010}, tendons \cite{Djordjevic2003}, cultured cartilage tissues \cite{Chen2004}, and cells \cite{Chen2004,Duenwald2009,Balland2006}. Fractional models such as the springpot model have been used to analyze the response in research on magnetic resonance elastography \cite{Klatt2010, Sack2013}. Fractional dynamics has become popular for modeling viscoelasticity, with experimental data and models reported for vessels \cite{Craiem2006}, the lung \cite{Suki1994, Yuan1997, Yuan2000}, skeletal muscle \cite{Klatt2010, Sack2013, GRAHOVAC2010, Kobayashi2012Soft, okamura2014study}, the brain \cite{Sack2009, Sack2013}, tendons \cite{Djordjevic2003}, the liver \cite{Sack2013,Kobayashi2005,Kobayashi2009,Kobayashi2012,kobayashi2012viscoelastic, Kobayashi2017}, breast tissues \cite{Tsukune2014,Kobayashi2012Enhanced}, muscle cells \cite{Chen2004}, blood cells \cite{Duenwald2009}, and living cells \cite{Balland2006}. 

The fractional dynamics model has been applied to a wide variety of materials, including biological materials. We previously developed a viscoelastic model based on the fractional dynamics model \cite{Kobayashi2005,Kobayashi2009,Kobayashi2012, kobayashi2012viscoelastic, Kobayashi2012Enhanced,Tsukune2014,Kobayashi2012Soft,okamura2014study, Kobayashi2017}.
The model was derived using experimental data obtained from in vitro measurements of a porcine liver \cite{Kobayashi2005,Kobayashi2009,Kobayashi2012,kobayashi2012viscoelastic}. We also validated the model with data obtained for in vitro breast tissue (mammary gland, fat, and muscle) \cite{Kobayashi2012Enhanced,Tsukune2014}. The model was partially  evaluated using data for in vitro and in vivo skeletal muscle tissue \cite{Kobayashi2012Soft, okamura2014study}.

Our previous study \cite{Kobayashi2017} also investigated the dynamic viscoelastic properties of liver tissue and evaluated the pairing of compliance \textit{$J(\omega)$} and phase delay \textit{$\phi(\omega)$}. 
The study showed that liver tissue has a power-law decrease in compliance  \textit{$J$}  and a constant phase delay \textit{$\phi$} in the frequency domain. These characteristics can be accurately represented using a fractional dynamics model. In the experiment and model, the compliance \textit{$J$} and phase delay  \textit{$\phi$} were found to be causally related via a non-integer derivative order \textit{$\alpha$}, specifically \textit{$J \propto \omega^{-\alpha}$}, \textit{$\phi = -\frac{\pi }{2}\alpha$}. In this case, the dynamic viscoelastic properties of liver tissue are represented by minimum phase properties \cite{das2011}. In previous studies, we also conducted a dynamic viscoelastic experiment on breast \cite{Kobayashi2012Enhanced,Tsukune2014} and muscle \cite{Kobayashi2012Soft} tissues. The results showed that the experimental data of breast and muscle tissues are not as highly correlated with the fractional model as the data for liver tissue are.
\subsection{\label{objectives} Objectives}

The objective of this study is to develop a fractional dynamics model that represents the viscoelastic properties of soft biological tissues. Specifically, a dynamic viscoelasticity test, which gives the frequency response, was conducted. We found that skeletal muscle and breast tissues have non-minimum phase properties; that is, the relationship between compliance and phase delay is not uniquely related to a non-integer derivative order \textit{$\alpha$}.  The experimental results show that the actual phase delay is larger than that estimated from compliance.

\begin{figure*}
\includegraphics[width=17cm]{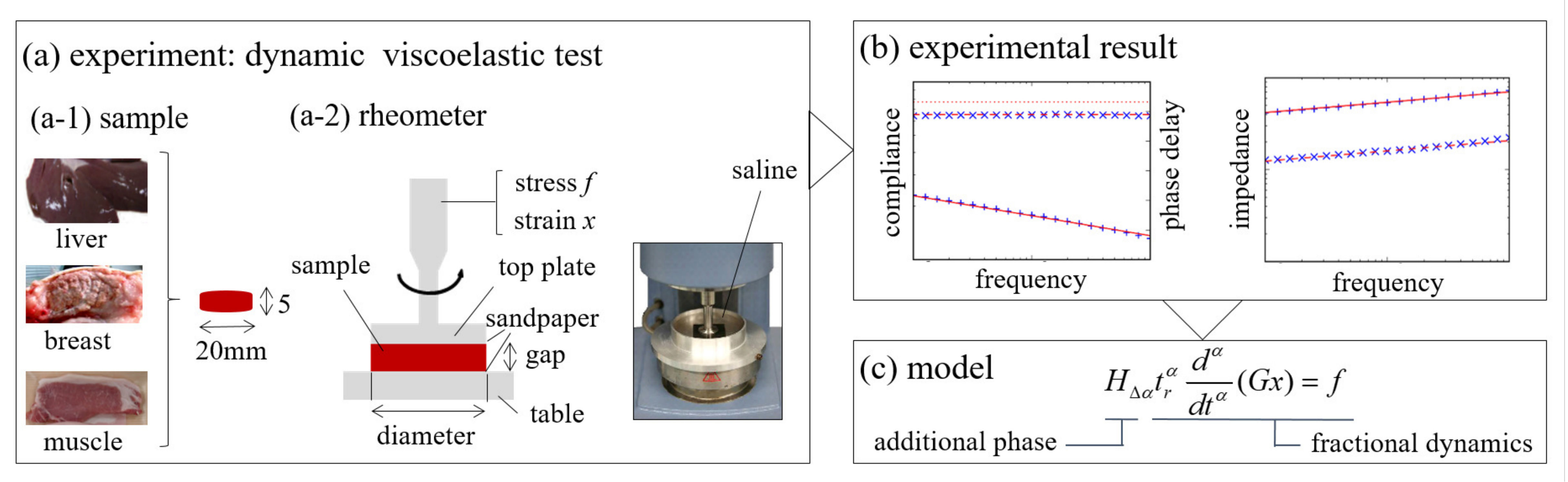}
\caption{ \label{fig:concept} 　
Visual overview of this article. The viscoelastic properties were investigated from material measurements of several biological tissues. We used porcine tissues (liver, mammary gland, breast muscle, breast fat, psoas major muscle, longissimus thoracis muscle, and muscle fat) as samples (a-1). We used a plate--plate rheometer, which can dynamically control and measure the stress and strain applied to the sample, to measure the samples (a-2). We conducted a dynamic viscoelastic test (b) to measure the viscoelastic properties and derive the viscoelastic model (c). We found that skeletal muscle and breast tissues have non-minimum phase properties; that is, the relationship between compliance and phase delay is not uniquely related to the fractional derivative order. The experimental results show that the actual phase delay is larger than that estimated from compliance. This paper also proposes a model for representing the non-minimum phase properties.
}
\end{figure*}

This paper proposes a model for representing the non-minimum phase properties obtained from a dynamic viscoelasticity test. We also show the amount of additional phase delay, defined as the increase in actual phase delay compared to that estimated from compliance, for several tissue types. Figure \ref{fig:concept} shows an overview of this article. 

\section{\label{mater}Materials and Methods}

\subsection{\label{proce} Materials}

We investigated the viscoelasticity of several types of porcine tissue, namely the liver, mammary gland, breast muscle, breast fat, and psoas major muscle, longissimus thoracis muscle, and muscle fat. For the liver \cite{Kobayashi2017} and breast \cite{Tsukune2014} tissues, we used the experimental data from a dynamic viscoelastic test reported in a previous study. We conducted an experiment on skeletal muscle tissues. Figure \ref{fig:concept} (a-2) shows the details of the measurement setup.

\subsection{\label{proce} Experimental setup and procedure}
The experimental setup and procedure are almost the same as those described in a previous article \cite{Kobayashi2017}. A description is given in this section to enhance the readability of this article.
 
We used a plate--plate rheometer (AR-G2 or DHR2; TA Instruments, New Castle, DE) to measure the stress and strain of the sample. A shear stress rheometer was selected because the shear test must be independent of any change in the cross-sectional area in the stress calculation. In addition, with this device, the effect of gravity can be disregarded. From these measurements, the conventional shear strain \textit{x} and conventional shear stress \textit{f} were calculated. The measurements of strain \textit{x} and stress \textit{f} are valid only when there is no slip between the sample and the plates. Thus, sandpaper was attached to the top plate and the measurement table to prevent sliding. The samples were cut into slices (diameter: 20 mm; thickness: about 5 mm), which were placed on a measurement table. The samples were soaked in a saline solution at 35$^\circ$C during testing.

After the saline solution had reached the target temperature, the gap between the table and the top plate was zeroed to the surface of the saucer. The saline solution was stable, and there was no reflux flow. Each tissue sample was placed on a measurement table, and the sample thickness (i.e., gap) was determined. The sample thickness was defined as the distance between the surface of the saucer and the surface of the parallel plate (part of the measurement device) at the time that the normal stress resulting from the contact between the parallel plate and the sample reached 0.1 N. To engage the sample and parallel plate, preloading for over 100 seconds and unloading for over 100 seconds were performed three times under a constant shear stress of 375 Pa. The following series of experiments were conducted for each sample after the above initialization procedures.

A sine-wave stress of 0.1 to 10 rad/s, providing a 1.5\% strain amplitude, was applied to the sample. The strain amplitude of 1.5\% (= 0.015) is within the range in which all tissues exhibited linear responses. The compliance \textit{J}, phase delay \textit{$\phi$}, storage elastic modulus \textit{G'}, and loss elastic modulus \textit{G''} at various angular frequencies $\omega$ were measured. Details of the process used to obtain the experimental results from the dynamic viscoelastic test are described in \cite{ Kobayashi2017}. The effects of the mass (inertia) and shear viscosity of the external normal saline solution could be disregarded at frequencies of lower than 10 rad/s. 
Data were collected for each tissue type. The number of samples for each tissue type is shown in Table \ref{tab:FundamentalStatistics}.
We obtained pairs of results, (compliance \textit{J}, phase \textit{$\phi$}) or (storage elastic modulus \textit{G'}, loss elastic modulus \textit{G''}, from the dynamic viscoelastic test.

\section{\label{resul}Results and Modeling}

\subsection{\label{bode} Compliance and phase delay}

\begin{figure*}
\includegraphics[width=17cm]{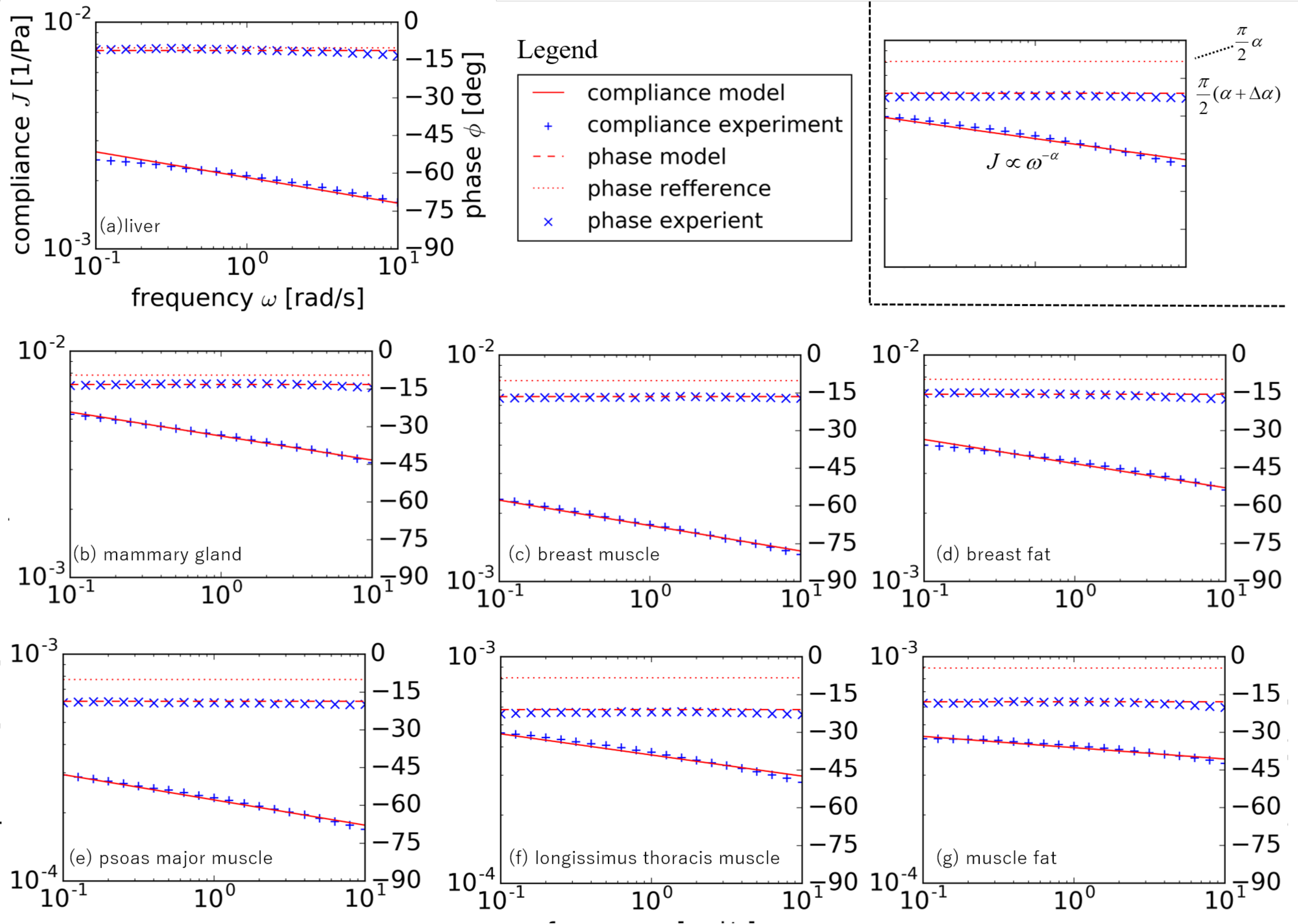}
\caption{\label{fig:BodeF} 
Compliance and phase delay diagrams. Typical experimental results for a sample of each tissue type are shown. Results for (a) liver, (b) mammary gland, (c) breast muscle,  (d) breast fat, (e) psoas major muscle (fillet), (f) longissimus thoracis muscle (loin), and (g) muscle fat. The plots show experimental data. All samples for a given tissue type exhibit the same trend as that of the typical sample. The power-law compliance \textit{J} decreases as the angular frequency \textit{$\omega$} increases for over two decades. The phase delay \textit{$\phi$} remains constant as the angular frequency \textit{$\omega$} changes for over two decades. The lines show the compliance \textit{J} and phase \textit{$\phi$}  of our model in equations (\ref{eq:GainLog}) and (\ref{eq:Phase}). The solid line shows the compliance \textit{J}. The dotted line shows the phase delay \textit{$\frac{\pi }{2} \alpha$} estimated from the compliance data. The dashed line shows the phase delay \textit{$\frac{\pi }{2} (\alpha+\Delta \alpha)$}. The difference between the dashed line and dotted line in the phase model is \textit{$\frac{\pi }{2} \Delta\alpha$}. For the liver tissue, there are a few differences between the dashed and dotted lines. Thus, the liver tissue has minimum phase viscoelastic properties. The other tissues have non-minimum phase viscoelastic properties. The model, which was fit to the typical experimental data through parameter identification, shows that the data of our model and the experimental data are highly correlated.
} 
\end{figure*}

Typical experimental results of the compliance and phase of a sample for each tissue type are shown in Fig.\ \ref{fig:BodeF}, where compliance \textit{J} is the multiplicative inverse of \textit{$G^*$}. The experimental data for all samples of a given tissue exhibited the same trend as that of the typical sample. The power-law compliance \textit{J} decreases as the angular frequency \textit{$\omega$} increases for over two decades. The phase delay \textit{$\phi$} remains constant as the angular frequency \textit{$\omega$} changes for over two decades.

The liver tissue response in the log-log diagram shown in Fig.\ \ref{fig:BodeF} (a) has almost the same slope as those for the mammary gland, breast muscle, breast fat, and psoas major muscle, shown in Fig.\ \ref{fig:BodeF} (b)-(e), respectively. This means that the power law index \textit{$\alpha$} from the compliance data is almost the same among these tissues. The phase delay in breast muscle, breast fat, and psoas major muscle is larger than that in liver tissue. The slope for longissimus thoracis muscle and muscle fat response, shown in Fig.\ \ref{fig:BodeF} (f)-(g) differs from the other tissues.
A model used in previous research on liver tissue \cite{Kobayashi2017} showed that the relationship between compliance (\textit{$J \propto \omega^{-\alpha}$}) and phase delay (\textit{$\phi== -\frac{\pi }{2} \alpha $}) is uniquely related to the derivative order \textit{$\alpha$}. The response of the liver tissue almost satisfies this relationship, but those of the other tissues do not. 
Thus, we found that some soft biological tissues have an additional phase delay, namely, the difference between the experimentally measured phase delay and the phase delay (\textit{$\phi= -\frac{\pi }{2} \alpha $}) estimated from the compliance data  (\textit{$J \propto \omega^{-\alpha}$}) . 

Here, we introduce a model that represents the characteristics of the experimental results, including the power-law form of compliance, constant phase delay, and additional phase delay. 
Our model is given in equation (\ref{eq:modelPresentAgain}). Equation (\ref{eq:modelPrevious}) is a model introduced in a previous article \cite{Kobayashi2017}; it is used as a reference. 

\begin{eqnarray}
{H_{\Delta \alpha }} t_r^\alpha \frac{{{d^\alpha }}}{{d{t^\alpha }}}(Gx) = f
\label{eq:modelPresentAgain}
\end{eqnarray}

\begin{eqnarray}
 t_r^\alpha \frac{{{d^\alpha }}}{{d{t^\alpha }}}(Gx) = f
\label{eq:modelPrevious}
\end{eqnarray}

where \textit{x} is the strain (torsional strain), \textit{f} is the stress (torsional stress), \textit{t} is time, \textit{$\alpha$} is a non-integer derivative order representing the viscoelasticity ratio, \textit{$t_r$} is the reference time scale, \textit{G} is the linear viscoelastic stiffness at an arbitrarily chosen point in time \textit{$t_r$}, \textit{${H_{\Delta \alpha }}$} is the fractional Hilbert transform operator of the order \textit{$\Delta \alpha$} \cite{lohmann1996fractional}, and \textit{$\Delta \alpha$} is an additional phase delay ratio used to represent the non-minimum amount of the system (i.e., an index of the additional phase delay). 
The term \textit{$H_{\Delta \alpha}$} is the fractional Hilbert transform operator, which is used to represent an additional phase delay. Equation (\ref{eq:modelPresentAgain}) is equal to equation (\ref{eq:modelPrevious}), presented in our previous study, when \textit{$\Delta \alpha=0$} \cite{lohmann1996fractional}.

The equation is expanded below to explain the above characteristics. The frequency transfer function is: 

\begin{eqnarray}
J(j\omega )= \frac{{X(j\omega)}}{{F(j\omega)}} = \frac{1} {j^{\Delta \alpha}} {t_r^\alpha G {{( {j \omega })}}^\alpha } = \frac{1}{j^{\Delta \alpha }{G{{\left( {j\frac{\omega }{{{\omega _r}}}} \right)}^\alpha }}}
\label{eq:Laplace2}
\end{eqnarray}

Here, \textit{$\omega$} is the angular frequency, \textit{j} is the imaginary unit, and $\omega_r$ is the reference scale, which is defined as \textit{$\omega_r = 1/t_r$}. We use the following relationship: \textit{$ H_{\Delta \alpha}=\exp ( j\frac{\pi }{2}\Delta \alpha )={j^{\Delta \alpha }}$}  \cite{lohmann1996fractional}.

The compliance \textit{J} is defined from equation (\ref{eq:Laplace2}) as follows:

\begin{eqnarray}
J(\omega ) = \left| {\frac{1}{{j^{\Delta \alpha }G{{(j\frac{\omega }{{{\omega _r}}})}^\alpha }}}} \right| = \frac{1}{{G{{\left( {\frac{\omega }{{{\omega _r}}}} \right)}^\alpha }}} = \frac{{J({\omega _r})}}{{{{\left( {\frac{\omega }{{{\omega _r}}}} \right)}^\alpha }}}
\label{eq:Gain}
\end{eqnarray}

where \textit{$J(\omega _r)$} is a coefficient representing compliance, which is defined as \textit{$J(\omega _r)=1/G$}.

Equation (\ref{eq:GainLog}) is derived from the log-log transformation of (\ref{eq:Gain}) through a transformation into dimensionless quantities.

\begin{eqnarray}
\log \left( {\frac{{J(\omega ){\kern 1pt} }}{{J({\omega _r})}}} \right)\,\; = {\kern 1pt} {\kern 1pt}  - \alpha \log \left( {\frac{\omega }{{{\omega _r}}}} \right)
\label{eq:GainLog}
\end{eqnarray}

The model equation of the phase delay \textit{$\phi$} is derived as follows:

\begin{eqnarray}
\begin{array}{l}
\phi \left( \omega  \right) = \arg \left( {\frac{1}{{{j^{\Delta \alpha }}G{{\left( {j\frac{\omega }{{{\omega _r}}}} \right)}^\alpha }}}} \right)\\
 =  - \arg \left( {{j^{\Delta \alpha }}} \right) - \arg \left( {G{{\left( {j\frac{\omega }{{{\omega _r}}}} \right)}^\alpha }} \right)\\
 =  - \frac{\pi }{2}\Delta \alpha  - \frac{\pi }{2}\alpha  =  - \frac{\pi }{2}(\alpha  + \Delta \alpha )\\
 = {\phi _o}
\end{array}
\label{eq:Phase}
\end{eqnarray}

where \textit{$\phi_o$ ($=-\frac{\pi }{2}(\alpha + \Delta \alpha))$} is the coefficient that represents the phase delay. 

Thus, our model represents the trends in the experimental results, namely the decrease in power-law compliance, as (\ref{eq:GainLog}), constant phase delay, and additional phase delay, as (\ref{eq:Phase}). 

We fitted the compliance \textit{$J(\omega)$} and phase \textit{$\phi(\omega)$} of our model to the experimental results through parameter identification. Specifically, the parameters \textit{$G$}, \textit{$\alpha$}, and \textit{$ \Delta \alpha$} were identified for each sample. Details of the method, process, and equation used in the parameter identification are provided in Appendix.

The data of compliance \textit{$J(\omega)$} and phase \textit{$\phi(\omega)$} from our model, in which the parameters were fitted to the experimental data, are shown in Fig.\ \ref{fig:BodeF}. The figure shows that the data of our model and the experimental data are strongly correlated. Table \ref{tab:FundamentalStatistics} lists the fundamental statistics about the model parameters for each tissue type. In Table \ref{tab:FundamentalStatistics}, each dataset for a single experiment was fitted to identify the set (\textit{$G$}, \textit{$\alpha$}, \textit{$\Delta \alpha$}) of model parameters. The results of these parameters were then averaged.

\subsection{\label{MI} Mechanical impedance}

In this section, we present the results of mechanical impedance. The mechanical complex impedance \textit{$G^*$} is defined as follows:

\begin{eqnarray}
{ {G^*(\omega)}  = G'(\omega) + jG''(\omega)}
\label{eq:defMI}
\end{eqnarray}

Here, $\omega$ is the angular frequency, \textit{$G^*$} is the complex mechanical impedance, \textit{G'} is the storage elastic modulus, and \textit{G''} is the loss elastic modulus. 

\begin{figure*}
\includegraphics[width=17cm]{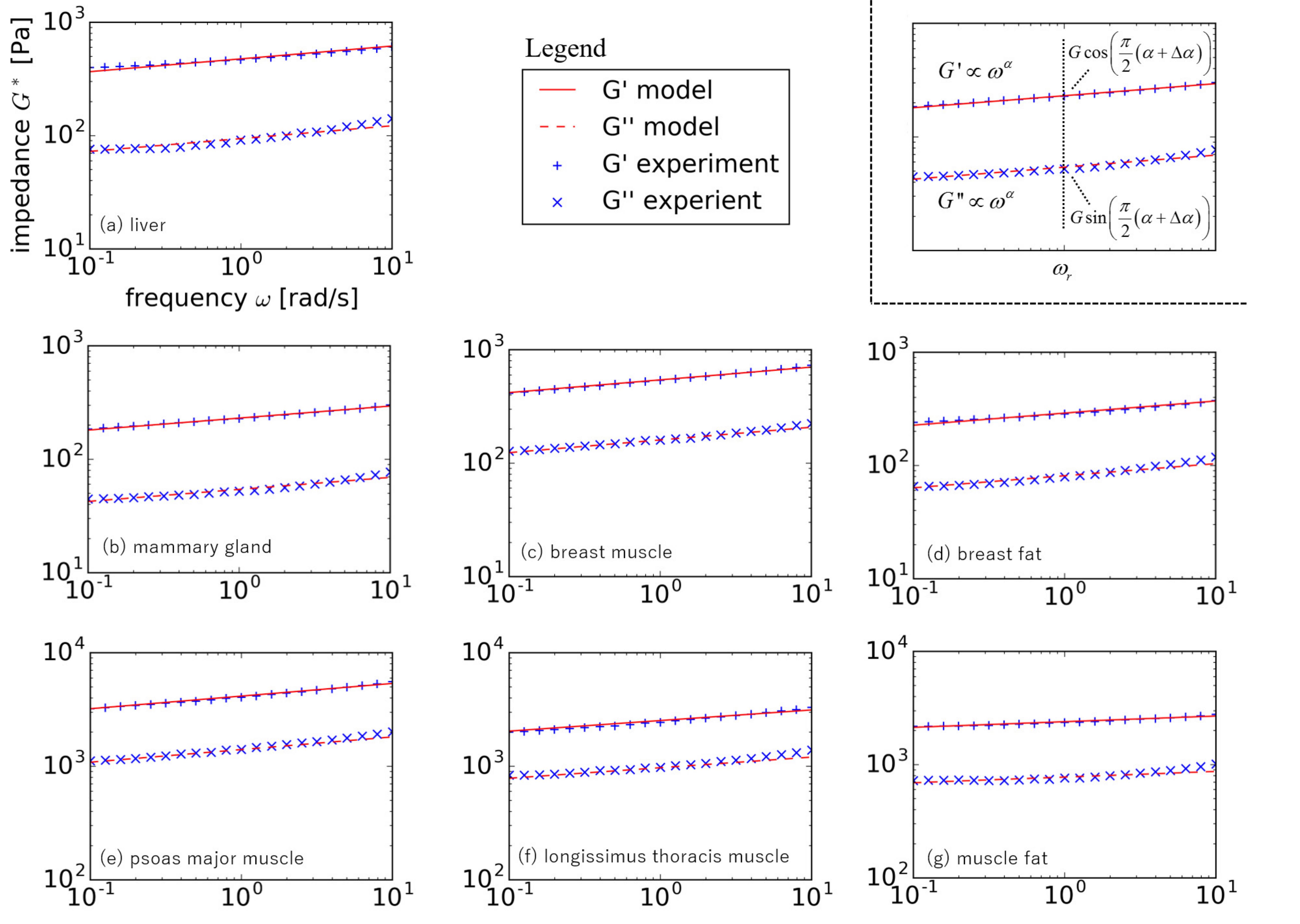}
\caption{\label{fig:MI} 
Mechanical complex impedance. Typical experimental data for the samples of each tissue type are shown. The plus and cross symbol plots respectively show the experimental data for the storage elastic modulus $G'$ and the loss elastic modulus \textit{$G''$}. Results for (a) liver, (b) mammary gland, (c) breast muscle, (d) breast fat, (e) psoas major muscle (fillet), (f) longissimus thoracis muscle (loin), and (g) muscle fat. All samples for a given tissue type exhibit the same trend as that of the typical sample.  \textit{$G'$} and \textit{$G''$} increase as the angular frequency \textit{$\omega$} increases. Both \textit{$G'$} and \textit{$G''$} exhibit a power-law form for over two decades. The slopes of \textit{$G'$} and \textit{$G''$} in the log-log diagram are almost the same. The data of \textit{$G'$} and \textit{$G''$} in our model are indicated by the solid and dashed lines, respectively. These data, which were fit to the typical experimental data through parameter identification, show that our model and the experimental data are highly correlated.
} 
\end{figure*}

Typical experimental data of the mechanical complex impedance \textit{$G^*$} for a sample of each tissue type are shown in Fig.\ \ref{fig:MI}. This figure was made using the data in Fig.\ \ref{fig:BodeF}. All samples for each tissue type exhibited the same trend as that of the typical sample. The storage elastic modulus \textit{$G'$} and the loss elastic modulus \textit{$G''$} increase with increasing angular frequency $\omega$. The data of \textit{$G'$} and \textit{$G''$} exhibit a power-law form for over two decades. The slopes of \textit{$G'$} and \textit{$G''$} in the log-log diagram are almost the same.

Our model shows the same characteristics as those of the experimental data, such as the power-law forms of  \textit{$G'$} and  \textit{$G''$}  with the same slopes. The equation is expanded below to explain the above results. Because equation (\ref{eq:modelPresentAgain}) takes the form of a frequency transfer function, the complex shear modulus \textit{$G^*$} can be expressed as follows:

\begin{eqnarray}
\begin{array}{l}
{G^*}(j\omega ) = \frac{{F(j \omega)}}{{X(j \omega)}} \\
=  j^{\Delta \alpha } G{\left( {j\frac{\omega }{{{\omega _r}}}} \right)^\alpha }= G{\left( {\frac{\omega }{{{\omega _r}}}} \right)^\alpha }{j^{(\alpha  + \Delta \alpha )}}\\

\end{array}
\label{eq:MIModel}
\end{eqnarray}
Here,  we use the following relationship: \textit{$ H_{\Delta \alpha}={j^{\Delta \alpha }}$}  \cite{lohmann1996fractional}. Equation (\ref{eq:MIModel}) expands to (\ref{eq:Gp}) and (\ref{eq:Gpp}) from (\ref{eq:defMI}) with a separation of the real and imaginary parts of (\ref{eq:MIModel}).

\begin{subequations}
\begin{eqnarray}
G'(\omega ) = G'({\omega _r}){\left( {\frac{\omega }{{{\omega _r}}}} \right)^\alpha }
\label{eq:Gp}
\\
G''(\omega ) = G''({\omega _r}){\left( {\frac{\omega }{{{\omega _r}}}} \right)^\alpha }
\label{eq:Gpp}
\end{eqnarray}
\label{eq:GpandGpp}
\end{subequations}

Here, \textit{$G'({\omega _r})$} and \textit{$G''({\omega _r})$} are constant parameters that represent the storage elastic modulus and the loss elastic modulus, respectively. The parameters have the following relationship (\ref{eq:GandGpandGpp2}):

\begin{subequations}
\begin{eqnarray}
G = \sqrt {G'{{({\omega _r})}^2} + G''{{({\omega _r})}^2}} 
\label{eq:Relation_Gp_Gpp}
\\
G'({\omega _r}) = G\cos (\frac{\pi }{2}(\alpha+\Delta \alpha) )
\label{eq:Gp2}
\\
G''({\omega _r}) = G\sin (\frac{\pi }{2}(\alpha+\Delta \alpha) )
\label{eq:Gpp2}
\end{eqnarray}
\label{eq:GandGpandGpp2}
\end{subequations}

Equations (\ref{eq:logGp}) and (\ref{eq:logGpp}) were derived from (\ref{eq:Gp}) and (\ref{eq:Gpp}) using a log-log transformation through a transformation into dimensionless quantities.

\begin{subequations}
\begin{eqnarray}
\log \left( {\frac{{G'(\omega )}}{{G'({\omega _r})}}} \right) = \alpha \log \left( {\frac{\omega }{{{\omega _r}}}} \right)
\label{eq:logGp}
\\
\log \left( {\frac{{G''(\omega )}}{{G''({\omega _r})}}} \right) = \alpha \log \left( {\frac{\omega }{{{\omega _r}}}} \right)
\label{eq:logGpp}
\end{eqnarray}
\end{subequations}

Thus, our model equation exhibits the same trend as that of the experimental data, i.e., the power-law dependence of the storage elastic modulus \textit{$G'$} and the loss elastic modulus \textit{$G''$}. The additional phase parameter \textit{$\Delta \alpha$} affects the ratio of storage elastic modulus \textit{$ G'(w_r)$} to loss elastic modulus \textit{$ G''(w_r)$}. This ratio for the model without the additional phase term (\ref{eq:modelPrevious}) is related to the power law index \textit{$\alpha$} as \textit{$ G'(w_r)/ G''(w_r) = tan(\frac{\pi }{2}\alpha)$}. 
This ratio for the model with the additional phase term (\ref{eq:modelPresentAgain}) is \textit{$ G'(w_r)/ G''(w_r) =tan(\frac{\pi }{2} (\alpha+ \Delta \alpha))$}.

The parameters \textit{$G$}, \textit{$\alpha$}, and \textit{$\Delta \alpha$} were identified by fitting the experimental data for all samples of each tissue type. The \textit{G'} and \textit{G''} in our model, which fit the typical experimental data, are presented in Fig.\ \ref{fig:MI}. This figure shows that the data of our model and the experimental data are strongly correlated. The coefficient of determination \textit{$R^2$} between our model and the experimental data for the series of \textit{G'} and \textit{G''} for all samples of each tissue type is approximately 90\%. Table \ref{tab:FundamentalStatistics} lists the fundamental statistics of the model parameters for each tissue type. In the table, each dataset for a single experiment was fitted to identify the set (\textit{$G$}, \textit{$\alpha$}, and \textit{$\Delta \alpha$}) of model parameters. The results of the model parameters were then averaged.

\begin{table*}
\caption{\label{tab:FundamentalStatistics}
Fundamental statistics of the model parameters for $t_r=1$ ($\omega_r=1$)}.
\begin{ruledtabular}
\begin{tabular}{ccccccccc}
\shortstack{tissue type}&
\shortstack{sample\\ number}&
\shortstack{\textit{G}\\ (Avg.)}&
\shortstack{\textit{G}\\ (S.D.)}&
\shortstack{\textit{$\alpha$}\\ (Avg.)}&
\shortstack{\textit{$\alpha$}\\ (S.D.)}&
\shortstack{\textit{$\Delta \alpha$} \\ (Avg.)}&
\shortstack{\textit{$\Delta \alpha$} \\ (S.D.)}&
\shortstack{\textit{$R^2$}\\ (Avg.)} \\
\hline
liver         &	6  & 402 &	132 &	0.120 & 0.008 & 0.003 &	0.011 & 0.90 \\
breast gland  & 10 & 252 & 63 &	0.111 & 0.010 &	0.042 &	0.008 &	0.91\\
breast muscle & 5　& 753 & 172&	0.116 &	0.003 & 0.068 &	0.013 & 0.91\\
breast fat    &	12 & 375 & 149&	0.107 & 0.007 &	0.067 & 0.017 & 0.92\\
psoas major muscle
              & 10 & 3586& 576& 0.114 & 0.011 &	0.092 & 0.014 & 0.93\\
longissimus thoracis muscle 
              & 10 &2738 & 462& 0.097 & 0.009 & 0.136 &	0.011 & 0.94\\
muscle fat
              & 10 &2006 & 471&	0.063 & 0.007 &	0.146 & 0.019 & 0.94 \\
\end{tabular}
\end{ruledtabular}
\end{table*}

\section{\label{discu}Discussion}
The main contribution of this article is the identification of the non-minimum phase viscoelastic properties of soft biological tissues and the development of a model that represents these properties. Here, minimum phase systems are defined as systems that have the minimum phase delay for a given magnitude (compliance in this article) of the response. A minimum phase system has the smallest possible phase for a give magnitude response. A system has minimum phase properties when it and its inverse are causal and stable. A non-minimum phase system has a phase delay that is larger than that of a minimum phase system with the equivalent magnitude. For a fractional-order system with index \textit{$\alpha$}, the system has minimum phase properties when  \textit{$\phi=\frac{\pi }{2} \alpha$}, and the system has non-minimum phase properties when \textit{$\phi>\frac{\pi }{2} \alpha$} \cite{das2011}. For a minimum phase system, the relationship between the magnitude (compliance) and phase delay is uniquely determined by Bode's theorem, which means that a phase diagram can be estimated from a magnitude diagram, and vice versa. For such a system, the time response can also be estimated from magnitude and phase diagrams through the inverse Fourier transform.

The index \textit{$\alpha$} in the fractional model for viscoelasticity is important for characterizing model properties. The value \textit{$\alpha$} can be estimated from several types of experimental data, such as a decrease in power-law compliance and constant phase delay. The estimation is not limited to the frequency domain. The time response, such as the power-law strain increase in the creep test and the power-law decrease in the stress relaxation test, can also be used. In this investigation, it was expected that the same value of \textit{$\alpha$} could be obtained in each experiment under the assumption that the above relationship in a minimum phase system is satisfied. The results obtained here show that the index \textit{$\alpha$} should be evaluated under the consideration that soft biological tissues have non-minimum phase viscoelastic properties. For example, we found a difference in the estimated index between the decrease in power-law compliance (\textit{$\alpha$}) and constant phase delay (\textit{$\alpha + \Delta\alpha$}). 

From a practical point of view, the contribution of this study is a parameter that is useful for discriminating tissue types. The additional phase delay parameter \textit{$\Delta \alpha$} differs with tissue type. In particular, muscle tissues such as the psoas major muscle, longissimus thoracis muscle, and muscle fat have very different \textit{$\Delta \alpha$} values. The \textit{$\Delta \alpha$} value may be related to the fat cell content in tissue. The liver has a simple cell structure and consists mainly of liver cells. Porcine liver tissue includes only a few fat cells, whereas breast and skeletal muscle tissues include many fat cells. In particular, the fat cell content in muscle tissue increases in the order of psoas major muscle, longissimus thoracis muscle, and muscle fat. The \textit{$\Delta \alpha$} value increases in the same order.

The main limitation of this study is that it does not explain how non-minimum phase properties come about. The Hilbert transform operator is used in the Benjamin-Ono equation for internal waves in stratified fluids, where it is introduced as a theoretical expansion of the physical model \cite{ono1975algebraic, benjamin1967internal}. Further theoretical investigation is needed regarding the fractional Hilbert transform and non-minimum phase properties. In addition, the effects of non-minimum phase properties on the response in the time domain should be investigated. Finally, the fractional model was partially explained through a fractal structure in related studies \cite{Kelly2009, Kobayashi2017}. The actual structure and how the non-minimum phase viscoelasticity and the fractional Hilbert transform operator can be related to the structure are still unknown.

\section{\label{conc}Conclusion}

This study proposed a model that represents the viscoelastic properties of soft biological tissues. We found that breast and skeletal muscle tissues have non-minimum phase properties in a dynamic viscoelastic test. The experimental results show that the actual phase delay is larger than the phase delay \textit{$\frac{\pi }{2} \alpha$} estimated from the index \textit{$\alpha$} of the power-law compliance. The proposed model and the experimental results were highly correlated in terms of the compliance and phase diagrams and the complex mechanical impedance. The additional phase delay parameter \textit{$\Delta \alpha$} may be useful for discriminating tissue types because it differs with tissue type.


\begin{acknowledgments}
This work was supported in part by the Japan Science and Technology Agency (JST) Precursory Research for Embryonic Science and Technology (PRESTO) (No. JPMJPR14D3), Japan, the Global Centers of Excellence (GCOE) Program and Grants for Excellent Graduate Schools, Japan, and a Grant-in-Aid for Scientific Research from the Ministry of Education, Culture, Sports, Science and Technology (MEXT) (No. 25350577), Japan.
\end{acknowledgments}

\appendix{

\section{\label{EKFforDynamicViscoelasticTest}Extended Kalman filter for dynamic viscoelastic test}

The parameter identification method was almost the same as that described in a previous article \cite{Kobayashi2017}. A description is given in this section to enhance the readability of this article.
This section shows the methodology used to identify the parameters described in Sec. \ref{resul}. The model for the dynamic viscoelastic test was as follows, derived from equations (\ref{eq:logGp})--(\ref{eq:logGpp}):

\begin{subequations}
\begin{eqnarray}
\log \left( {\frac{{G'(\omega )}}{{G'({\omega _r})}}} \right) = \alpha \log \left( {\frac{\omega }{{{\omega _r}}}} \right)
\label{eq:AppC1a}
\\
\log \left( {\frac{{G''(\omega )}}{{G''({\omega _r})}}} \right) = \alpha \log \left( {\frac{\omega }{{{\omega _r}}}} \right)
\label{eq:AppC1b}
\\G'({\omega _r}) = G\cos (\frac{\pi }{2}　(\alpha + \Delta \alpha) )
\label{eq:AppC1c} 
\\
G''({\omega _r}) = G\sin (\frac{\pi }{2} (\alpha + \Delta \alpha) )
\label{eq:AppC1d}
\end{eqnarray}
\end{subequations}

where \textit{G'}, \textit{G''}, and \textit{$\omega$} are variables, and \textit{G}, \textit{$\alpha$}, and \textit{$\Delta \alpha$} are parameters. 

We obtained the set of \textit{G'} and \textit{G''} at each angular frequency \textit{$\omega$} value from the experiment. We identified the parameterfrom these data using the extended Kalman filter (EKF) (ref. \cite{Hoshi2008}). System identification using the EKF can be generally described as follows: 

\begin{subequations}
\begin{eqnarray}
{\theta _{k + 1}} = f({\theta _k},{\psi _k})
\label{eq:AppC2a} 
\\
{y_k} = g({\theta _k},{\zeta _k})
\label{eq:AppC2b} 
\end{eqnarray}
\end{subequations}
where \textit{k = 0, 1, 2,...} represents the discrete iteration index (number of datasets in this case), \textit{$\theta$} is an n-dimensional state vector, \textit{$\psi$} is an n-dimensional system noise vector, \textit{y }is a p-dimensional observation vector, \textit{$\zeta$} is a p-dimensional observation noise vector, and \textit{f()} and \textit{g()} are nonlinear vector functions. In state-space theory, (\ref{eq:AppC2a}) and (\ref{eq:AppC2b}) are known as the system model (or state model) and the observation model, respectively. 

The parameter vector is regarded as a state vector in the EKF for system identification. The state vector (parameter vector) \textit{$\theta$} is a constant vector and the observation noise vector \textit{$\zeta$} is a Gaussian white noise with zero mean. (\ref{eq:AppC2a}) and (\ref{eq:AppC2b}) are represented as:

\begin{subequations}
\begin{eqnarray}
{\theta _{k + 1}} = I{\theta _k}
\label{eq:AppC3a} 
\\
{y_k} = h({\theta _k}) + {\zeta _k}
\label{eq:AppC3b} 
\end{eqnarray}
\end{subequations}
where \textit{I} is the identity matrix and \textit{h()} is a nonlinear vector function. For system identification for the dynamic viscoelastic test, the state vector (parameter vector) \textit{$\theta$}, observation vector \textit{y}, and nonlinear vector function \textit{h()} are regarded as follows for \textit{$\omega_r=1$}: 

\begin{subequations}
\begin{eqnarray}
\theta  &=& \left[ {\begin{array}{*{20}{c}}G \\ \alpha\\ \Delta \alpha \end{array}} \right]
\label{eq:AppC4a} 
\\
y &=& \left[ {\begin{array}{*{20}{c}}
{\log G'}\\
{\log G''}
\end{array}} \right]
\label{eq:AppC4b} 
\\
h(\theta ) &=& \left[ {\begin{array}{*{20}{c}}
{\alpha \log \omega + \log (G\cos (\frac{\pi }{2}(\alpha + \Delta \alpha )) )}\\
{\alpha \log \omega + \log (G\sin (\frac{\pi }{2}(\alpha + \Delta \alpha )))}
\end{array}} \right]
\label{eq:AppC4c} 
\end{eqnarray}
\end{subequations}

The EKF algorithm (ref.\ \cite{Hoshi2008}) using (\ref{eq:AppC4a})--(\ref{eq:AppC4c}) was applied to identify the parameter from the dataset. It was not necessary to set initial values for each parameter \textit{$\theta_0$}, meaning that \textit{$\theta_0$} was a zero vector.     

｝


\bibliographystyle{apsrev} 

\bibliography{Manuscript}

\end{document}